\documentstyle[12pt]{article}
\begin{document}

\bibliographystyle{unsrt}
\def\Journal#1#2#3#4{{#1} {\bf #2}, #3 (#4)}
\def\NCA{ Nuovo Cimento}
\def\NIM{ Nucl. Instrum. Methods}
\def\NIMA{{ Nucl. Instrum. Methods} A}
\def\NP{{ Nucl. Phys.} }
\def\PLB{{ Phys. Lett.}  B}
\def\PRL{ Phys. Rev. Lett.}
\def\PRD{{ Phys. Rev.} D}
\def\ZPC{{Z. Phys.} C}
\def\ZPA{{Z. Phys.} A}
\def\st{\scriptstyle}
\def\sst{\scriptscriptstyle}
\def\mco{\multicolumn}
\def\epp{\epsilon^{\prime}}
\def\vep{\varepsilon}
\def\ra{\rightarrow}
\def\ppg{\pi^+\pi^-\gamma}
\def\vp{{\bf p}}
\def\ko{K^0}
\def\kb{\bar{K^0}}
\def\al{\alpha}
\def\ab{\bar{\alpha}}
\def\be{\begin{equation}}
\def\ee{\end{equation}}
\def\bea{\begin{eqnarray}}
\def\eea{\end{eqnarray}}
\def\CPbar{\hbox{{\rm CP}\hskip-1.80em{/}}}
\def\ua{\uparrow}
\def\da{\downarrow}


\section*{Perturbative QCD study on the 
photonic penguin contributions to the decay $B\rightarrow K^* \gamma$}

\vskip 0.5cm
\centerline{Fuguang Cao$^{a,b,c}$ and Tao Huang$^{a,b}$}

\vskip 0.5cm
\centerline{$^a$ CCAST (World Laboratory), P.O. Box 8730,
Beijing 100080, P.~R.~China}
\centerline{$^b$ Institute of High Energy Physics, Academia Sinica,
P.O. Box 918, Beijing 100039, P.~R.~ China\footnote{Mailing address.
E-mail address: caofg@itp.ac.cn}}
\centerline{$^c$ Institute of Theoretical Physics, Academia Sinica,
P.O. Box 2735, Beijing 100080, P.~R.~China}

\vskip 2cm
\begin{abstract}
Including corrections of order $O(m_{K^*}/m_B)$,
we present an analysis of photonic penguin contributions
to the decay $B\rightarrow K^* \gamma$ in the perturbative QCD framework.
Employing several models of the meson wave functions,
we demonstrate that the corrections of $O(m_{K^*}/m_B)$ 
are enhanced and will provide substantial contributions to the decay
because of the $B$ meson wave function being sharply peaked
(bound state effect).
The numerical predictions for the corrections are about $30\%\sim 60\%$
which depend on the non-perturbative inputs such as the meson wave 
functions and the $b$-quark mass.
\end{abstract}


\newpage
\noindent
{\bf 1. Introduction}
\smallskip

The rare decay $B\rightarrow K^* \gamma$ has attracted great attentions
especially after the CLEO Collaboration first identified this decay and
gave its branching ratio \cite{CLEO}.
The decay $B\rightarrow K^* \gamma$ is dominated by the flavor-changing
quark-level process $b \rightarrow s \gamma$ which can occur not only
through penguin diagram at one-loop level in the standard model (SM) but also
through virtual particle in the supersymmetry and other extensions of
the standard model \cite{NSM1,NSM2}.
Thus accurate experimental measurements and theoretical calculations
of this decay can provide a precision test of the standard model 
as well as a test of new physics at present experimentally accessible
energy scale.
It has been pointed out \cite{Brod} that perturbative QCD (PQCD) may be
applicable to the exclusive nonleptonic decays of $B$ meson  since
there is a hard-gluon exchange between the heavy and light quarks
in these decays.
Recently calculations also show that PQCD may give a good description of the 
two body hadronic decays of $B$ meson \cite{Carlson}.

In the standard model (SM),
the mainly contribution to the decay $B\rightarrow K^* \gamma$
comes from the photonic penguin diagrams which are shown in Fig.~1.
Compared to Fig.1a, the contribution from Fig.~1b is of order $m_{K^*}/m_B$,
and thereby it is not included in Ref. \cite{CarlsonK}.
In this paper we shall re-analyse decay $B\rightarrow K^* \gamma$ in the SM
by including $O(m_{K^*}/m_B)$ corrections in the amplitude.
Including bound state effect and employing several models of distribution
amplitudes of $B$ and $K^*$ mesons, we find that these corrections 
are enhanced by the bound state effect and become more important.
This paper is organized as follows: In section 2, 
we calculate the photonic penguin diagram contributions to the order
$m_{K^*}/m_B$ in the amplitude after describing the effective Hamiltonian.
In section 3, we present numerical results by employing several
models of meson distribution amplitudes.
As usual, the last section is reserved for summary.

\bigskip
\noindent
{\bf 2. Contribution coming from photonic penguin diagram}
\smallskip

The effective Hamiltonian (the square blob part in Fig.~1)
which describes the photonic penguin diagram,
can be expressed as \cite{Heff1,Heff2,Heff3}
\begin{eqnarray}
H_{eff}=-4 \frac{G_F}{\sqrt{2}} V_{tb}V^*_{ts} C_7(\mu) O_7(\mu),
\end{eqnarray}
where
\begin{eqnarray}
O_7(\mu)=\frac{e}{16 \pi^2}m_b \bar s \sigma^{\mu\nu} F_{\mu\nu}
\frac{1}{2}(1+\gamma_5)b.
\end{eqnarray}
In the above expressions, $C_7(\mu)$ is the Wilson coefficient which
contains the effects of QCD corrections,
\begin{eqnarray}
C_7(\mu)=\eta^{-16/3\beta_0}\left[C_7(m_W)-\frac{58}{135}
\left(\eta^{10/3\beta_0}-1 \right)-\frac{29}{189}
\left(\eta^{28/3\beta_0}-1 \right)\right],
\end{eqnarray}
where $\eta=\alpha_s(\mu)/\alpha_s(m_W)$, $\beta_0=11-(2/3)n_f$ and
$C_7(m_W)=-0.19$ is given in the $W$-mass scale.

The wave function of the $B$ meson can be written in the form \cite{Brod}
\begin{eqnarray}
\psi_B=\frac{1}{2}\frac{I_c}{\sqrt 3}\phi_B(x)\gamma_5(\rlap/ p_B -m_B),
\end{eqnarray}
where $I_c$ is the identity in the color space.
For the $K^*$ meson, the wave function can be expressed as
\begin{eqnarray}
\psi_{K^*}=\frac{1}{2}\frac{I_c}{\sqrt 3}\phi_{K^*}(x)
\rlap/ \xi^*(\rlap/ p_{K^*}+m_{K^*}),
\end{eqnarray}
where $\xi^*$ is the polarization vector of the $K^*$ meson.
$\phi_B$ and $\phi_{K^*}$ are the distribution
amplitudes of the $B$ and $K^*$ meson respectively.

We express the contribution to the amplitude
in the gauge invariant form\footnote{It is 
worthwhile to note that the expression for the amplitude presented here,
Eq. (\ref{MT}), is gauge invariant, while the one given in \cite{CarlsonK}
is not because of the second term in the bracket of Eq. (\ref{MT})
being missed \cite{Zhang}.}
\begin{eqnarray}
M_i=t_i\times \frac{1}{2 p_B\cdot q}\left[p_B\cdot q \epsilon^* \cdot \xi^*
-p_B\cdot\epsilon^* q\cdot\xi^*
+i\epsilon_{\mu \nu \alpha \beta} p_B^\mu q^\nu \epsilon^{*\alpha}
\xi^{*\beta}\right].
\label{MT}
\end{eqnarray}
The contributions from Figs. 1a and 1b can be written as
\begin{eqnarray}
t_1&=&\displaystyle{G m_b\int [dx][dy]\phi_B(x)\phi_{K^*}(y)
\frac{1}{l_b^2-m_b^2}\frac{1}{k_g^2}} \nonumber \\
& &\times Tr \left\{\frac{\gamma_5(\rlap/ p_B-m_B)}{\sqrt 2}\gamma^\alpha
\frac{\rlap/ \xi^*(\rlap/ p_{K^*}+m_{K^*})}{\sqrt 2}
\sigma^{\mu\nu}F_{\mu\nu}\frac{1}{2}(1+\gamma_5)(\rlap/ l_b+m_b)
\gamma_\alpha \right\} \nonumber \\
&=& 4Gm_b \int_0^1 [dx]\frac{1}{x_1}\phi_B(x)
   \int_0^1 [dy] \frac{(1-y_1)m_B^2-2 m_b m_B }
   {y_1 \left[m_b^2-(1-y_1)m_B^2\right]}\phi_{K^*}(y) \nonumber \\
& &+ 4Gm_b \int_0^1 [dx]\frac{1}{x_1}\phi_B(x) 
   \int_0^1 [dy]\frac{\left[m_b-2(1-y_1)m_B\right]m_{K^*} }
{y_1 \left[m_b^2-(1-y_1)m_B^2\right]}\phi_{K^*}(y) \nonumber \\
&\equiv & 4Gm_bI_{1B}I_{1K^*}^{LO} + 
4Gm_bI_{1B}I_{1K^*}^{NLO} \nonumber \\
&\equiv &t_1^{LO}+t_1^{NLO}
\label{t1}
\end{eqnarray}
and
\begin{eqnarray}
t_2^{NLO}&=&\displaystyle{G m_b\int [dx][dy]\phi_B(x)\phi_{K^*}(y)
\frac{1}{l_b^2-m_b^2}\frac{1}{k_g^2}} \nonumber \\
& &\times Tr \left\{\frac{\gamma_5(\rlap/ p_B-m_B)}{\sqrt 2}\gamma^\alpha
\frac{\rlap/ \xi^*(\rlap/ p_{K^*}+m_{K^*})}{\sqrt 2}
(\rlap/ l_b+m_b)\sigma^{\mu\nu}F_{\mu\nu}\frac{1}{2}(1+\gamma_5)
\gamma_\alpha \right\} \nonumber \\
&=& -4 Gm_b \int_0^1 [dx]\frac{1-x_1}{x_1^2}\phi_B(x)
\int_0^1 [dy]\frac{m_{K^*}}{y_1 m_B}\phi_{K^*}(y) \nonumber \\
&\equiv&-4Gm_b I_{2B}I_{2K^*}^{NLO},
\label{t2}
\end{eqnarray}
In the above expressions, $[dx]=dx_1dx_2 \delta(1-x_1-x_2)$,
$[dy]=dy_1dy_2 \delta(1-y_1-y_2)$, $q$ and $\epsilon$ are
the momentum and polarization of the photon respectively, and 
\begin{eqnarray}
G=\frac{G_F}{\sqrt 2 \pi} V_{tb}V_{ts}^*C_F C_7(\mu)e\alpha_s(\mu).
\end{eqnarray}
$x_1$ and $y_1$ in Eqs. (\ref{t1}) and (\ref{t2})
are the momentum fractions carried by the
light quarks in the $B$ and $K^*$ mesons respectively.
The distribution amplitude of $B$ meson,
$\phi_B(x)$, should be sharply peaked
at some small value of $x_1$ since $m_b$ is much larger
than the light quark mass \cite{Brod,CarlsonK}.
Thus we keep only the leading contributions of $x_1$ in the 
quark and gluon propagators,
which are $x_1$ terms in Fig.~1a and $x_1^2$ terms in Fig.~1b.
The fermion propagators in Figs.~1a and 1b contribute different
factors to $t_1$ and $t_2$:
The one in Fig.~1a involving only $K^*$ meson variable
in the form of $1/[y_1 m_b^2-(m_B^2-m_b^2)]$ is attributed to
the integrals $I_{1K^*}^{LO}$ and $I_{1K^*}^{NLO}$;
The one in Fig.~1b involving only $B$ meson variable 
in the form of $1/(x_1 m_B^2)$ is attributed to
the integral $I_{2B}$. 
The gluon propagators in Figs.~1a and 1b involving both 
$B$ and $K^*$  meson variables
in the form of $1/(x_1 y_1 m_B^2)$ can be factored to the  
integrals $I_{iB}$ and $I_{iK}$.
In this way, $t_i$ is factorized to two independent integrals
$I_{iB}$ and $I_{iK^*}$.

In Eqs. (\ref{t1}) and (\ref{t2}), $t_1^{LO}$ provides leading
contribution while $t_1^{NLO}$ and $t_2^{NLO}$ are corrections
of $O(m_{K^*}/m_B)$. 
It is interested to notice that the suppression factor $m_{K^*}/m_B$ 
in $t_2^{NLO}$ can be compensated by the bound state effect as
it is going to be demonstrated in the following.
Compared to $I_{1B}$, the fermion propagator in
Fig.~1b provides an additional factor $1/x_1$ to $I_{2B}$.
Because the distribution amplitude of $B$ meson, $\phi_B$, is
sharply peaked at $x_1\approx 0.05\sim 0.1$ \cite{Brod}, 
$I_{2B}$ is much larger than $I_{1B}$.
For example, employing a simple model for $\phi_B$,
$\phi_B\sim \delta(x_1-\epsilon_B)$ with
\bea 
\epsilon_B=\frac{m_B-m_b}{m_B},  
\label{epsilonB}
\eea
the ratio is (see Table 1)
\begin{eqnarray}
\frac{I_{2B}}{I_{1B}}=\frac{1-\epsilon_B}{\epsilon_B}
=\frac{m_b}{m_B-m_b} \approx 10 \sim 18.
\end{eqnarray}
This factor will cancels approximately the suppression factor
$m_{K^*}/m_{B}$ being about $1/17$ in $I_{2K^*}^{NLO}$,
which make the contribution coming from Fig.~1b become important.
There is no similar enhancement factor in $t_1^{NLO}$,
so it is order $m_{K^*}/m_{B}$ and may be neglected as compared to $t_1^{LO}$.

It has been  pointed out \cite{CarlsonK} that the contribution
coming from Fig.~1a, $t_1^{LO}$, contains a large imaginary part 
because of the pole of the heavy quark propagator.
This imaginary part does not correspond to the long-distance physics.
It should be noticed that the ratio of the imaginary part to the real part 
depends on the $b$-quark mass $m_b$ (namely $\epsilon_B$)
and $B(K^*)$ distribution amplitudes (see Table 2), which are about
$3.5 \sim 0.8$. Thus the contribution form Fig. 1b should be
taken into account although it provides only a real contribution.

The decay width and branching ratio can be obtained readily,
\begin{eqnarray}
\Gamma =\frac{1}{16 \pi m_B}\left|\sum_{polarization} (M_1+M_2) \right|^2,
\end{eqnarray}
\begin{eqnarray}
Br(B\rightarrow K^* \gamma)=\frac{\Gamma}{\Gamma_{total}}.
\end{eqnarray}

\bigskip
\noindent
{\bf 3. Numerical calculation and model analysis}
\smallskip

For the numerical results, we take the following parameters as inputs:
\begin{eqnarray}
\Lambda_{QCD} = 200\,\, {\rm MeV}, & \mu = 1 \,\,{\rm GeV}, \nonumber \\
m_W = 81\,\, {\rm GeV}, & m_t = 2 m_W, \nonumber \\
V_{tb}= 0.999, & V_{ts} = -0.045, \\
f_B= 132\,\, {\rm MeV} \cite{fB}, & ~~~~f_{K^*} = 151 \,\,
{\rm MeV}\cite{fK},\nonumber\\
\tau_B = 1.46\times 10^{-12}\,\, {\rm second}. \nonumber 
\end{eqnarray}

The numerical results should depend on the expressions of distribution
amplitudes $\phi_B(x)$ and $\phi_{K^*}(y)$ which are determined by 
the non-perturbative physics.
For the $\phi_{B}$ we adopt the following models:
i) According to Brodsky-Huang-Lepage prescription \cite{BHL}
the $B$ meson wave function can be given in the form \cite{Guo},
\begin{eqnarray}
\psi_B(x,k_\bot)=A {\rm exp}
\left[-b^2 \left(\frac{m_b^2+k_\bot^2}{x_2}+\frac{m_q^2+k_\bot^2}{x_1}
\right)\right],
\end{eqnarray}
in which the parameters $A$ and $b$ are determined by two constraints:
\begin{eqnarray}
\int_0^1 [dx] \frac{d^2 {\bf k}_\bot}{16\pi^3} \psi_B(x,k_\bot) 
= \frac{f_B}{2 \sqrt 3},
\label{constraint1}
\end{eqnarray}
and 
\begin{eqnarray}
P_B = \int_0^1 [dx]\frac{d^2  {\bf k}_\bot}{16 \pi^3}
\left|\psi_B(x,k_\bot)\right|^2\approx 1.
\label{constraint2}
\end{eqnarray}
$P_B$ is the probability of finding the $|q \bar q \rangle$ Fock state in the 
$B$ meson. The second constraint $P_B\approx 1$ is reasonable since with
the increase of the constitute quark mass the valence Fock state occupies
the most fraction in the hadron, and in the nonrelativistic limit the
probability of finding the valence Fock state is going to approach unity.
Then we can obtain the distribution amplitude of $B$ meson
\begin{eqnarray}
\phi_B^{BHL}(x)=\frac{A}{16\pi^2 b^2} x_1 x_2 {\rm exp}
\left[-b^2\left(\frac{m_b^2}{x_2}+\frac{m_q^2}{x_1}\right)\right].
\end{eqnarray}
ii) Szczepaniak, Henley and Brodsky suggested another model
for $\phi_B(x)$ \cite{Brod},
\begin{eqnarray}
\phi_B^{SHB}(x)=\frac{A}{\left(\epsilon_B^2/x_1+1/x_2 -1\right)^2},
\end{eqnarray}
where $A$ and $\epsilon_B$ are given by Eqs. (\ref{constraint1}) and 
(\ref{epsilonB}) respectively.
iii) The simplest model for $\phi_{B}$ is the $\delta$-function
approximation which has been adopted in Refs. \cite{Carlson,CarlsonK}
\begin{eqnarray}
\phi_B^{\delta}(x)=\frac{f_B}{2\sqrt 3}\delta(x_1-\epsilon_B),
\label{waveCarl}
\end{eqnarray}
where $\epsilon_B$ is related to the longitudinal momentum fraction 
of the light quark (see Eq. (\ref{epsilonB})).

We adopt the following two models for $\phi_{K^*}$:
i) it has been pointed out \cite{Guo,Xiang,ZD} 
that $K^*$ meson wave function is close to its asymptotic behavior,
so we adopt the expression in Ref. \cite{Guo},
\begin{eqnarray}
\phi_{K^*}(x)=\frac{A}{16\pi^2 b^2} y_1 y_2 {\rm exp}
\left[-b^2\left(\frac{m_s^2}{y_2}+\frac{m_q^2}{y_1}\right)\right],
\label{Kwf}
\end{eqnarray}
where $A=41.4\,\,{\rm GeV}^{-1}$, $b=0.74\,\,{\rm GeV}^{-1}$,
$m_s=0.55\,\,{\rm GeV}$ and $m_q=0.35\,\, {\rm GeV}$.
The quark masses appearing in the meson wave functions
(distribution amplitudes) should be the constituent quark masses
since the wave function is determined mainly by the soft-physics,
while the quark masses appearing in the hard amplitudes should
be the current quark masses which can be ignored reasonably for the
light quarks.
ii) The asymptotic expression for $\phi_{K^*}$,
\bea
\phi_{K^*}(y)=\sqrt{3}f_{K^*}y_1 y_2.
\eea

The numerical results are given in tables~ 2 and 3.
$\phi_B^{BHL}$, $\phi_B^{SHB}$ and $\phi_B^{\delta}$ 
have different behavior in the $x$-space (see Fig.~2).
$\phi_B^{BHL}$ is not so sharply peaked as
$\phi_B^{SHB}$ and $\phi_B^{\delta}$,
and the position of the maximum of $\phi_B^{BHL}$ is farther away 
the end-point $x_1=0$ than that of the other two models 
{\sl i.e.} $\phi_B^{BHL}$ does not emphasize the small-$x_1$ region 
so strongly as $\phi_B^{SHB}$ and $\phi_B^{\delta}$.
Thus the value of $I_{1B}$ ($I_{2B}$) calculated with $\phi_B^{BHL}$
is the smallest one among the three models (see table~1).
Because of $t_1$ and $t_2$ depending on $\phi_B$ only through the
integrals $I_{1B}$ and $I_{2B}$ respectively (see Eqs. (\ref{t1}) and
(\ref{t2})), the decay amplitude and
branching ratio calculated with $\phi_B^{BHL}$ will be also the 
smallest one  (see table~3).

It can be found that $t_1^{NLO}$ is about $1/10$ of $t_1^{LO}$ because 
of the suppression factor $m_{K^*}/m_B$, 
while $t_2^{NLO}$ is the same order as the real part of $t_1^{LO}$ since
the bound state effect compensates approximately
the suppression factor $m_{K^*}/m_B$ (see table~2).
The corrections to the decay branching ratio are about $30\% \sim 60\%$
which varies with the distribution amplitudes of $B$ and $K^*$
mesons and the $b$-quark mass.
The corrections calculated with $\phi_B^{SHB}$ 
is more important than that with the other two models,
and the corrections calculated with $\phi_{K^*}^{GH}$ and $\phi_{K^*}^{as}$ 
are very similar since they have similar behavior.
It can been found also that the corrections become more important 
with $m_b$ increasing.
We would like to point out again that 
it is because the distribution amplitude of $B$ meson
should be sharply peaked at some small value of $x_1=\epsilon_B$
(the bound state effect)
that the corrections of $O(m_{K^*}/m_B)$ coming from Fig.~1b
become more important.
 
As comparing with the experimental data, we find that
the results calculated with $\phi_B^{SHB}$ and $\phi_B^{\delta}$,
and with $m_b$ being about $4.9$
are comparable to the experiment data 
$Br(B\rightarrow K^* \gamma)=4.5\pm 1.5 \pm 0.9\times 10^{-5}$ 
\cite{CLEO}.

\bigskip
\noindent
{\bf 4. Summary}
\smallskip

The decay $B \rightarrow K^* \gamma$ is a very attractive process
since it provides an experimentally accessible way for 
a subtle test of the standard model and a test of new physics.
Both more accurate
theoretical calculations and more accurate experimental measurements
about this decay mode are worthwhile and necessary.
By including the corrections of order $m_{K^*}/m_B$ in the
photonic penguin diagrams, we have analysed the decay
$B \rightarrow K^* \gamma$ in the framework of perturbative QCD.
Employing several models of the meson wave functions,
we find that the $O(m_{K^*}/m_B)$ corrections coming from Fig.~1b 
provides substantial corrections to the branching ratio
since the bound state effect provides an enhancement factor 
$m_b/(m_B-m_b)$ which cancels approximately the suppression
factor $m_{K^*}/m_B$.
The corrections are about about $30\%\sim 60\%$
which depend on the non-perturbative inputs such as the meson wave 
functions and the $b$-quark mass.

\bigskip
\noindent
{\sl Acknowledgements:}
F. G. Cao would like to thank J. Cao, D. X. Zhang and Y. D. Yang 
for helpful discussions. This work partially supported
by the Postdoc Science Foundation of China.

\newpage

\newpage
\begin{center}
\begin{tabular}{|c|c|c|c|c|c|c|c|c|c|} \hline
& \multicolumn{3}{|c|}{$\phi_B^{BHL}$} & 
\multicolumn{3}{|c|}{$\phi_B^{SHB}$} &
\multicolumn{3}{|c|}{$\phi_B^{\delta}$} \\ \hline
$m_b({\rm GeV})$ & 4.8 & 4.9 & 5.0 & 4.8 & 4.9 & 5.0 & 4.8 & 4.9 & 5.0 \\\hline
$I_{1B}$& 0.29 & 0.30 & 0.31 & 0.38 & 0.46 & 0.59 & 0.42 & 0.53 & 0.72 \\ \hline
$I_{2B}$& 3.50 & 3.63 & 3.77 & 7.16 & 10.7 & 18.4 & 4.20 & 6.83 & 12.8 \\ \hline
\end{tabular}
\vskip 0.5cm
Table 1. $\phi_B$-dependence of $t_i$.
\end{center}

\vskip 1.cm
\begin{center}
\begin{tabular}{|c|c|c|c|c|c|c|c|} \hline
\multicolumn{2}{|c|}{} & 
\multicolumn{3}{|c|}{$\phi_{k^*}^{GH}$} & 
\multicolumn{3}{|c|}{$\phi_{k^*}^{as}$} \\ \hline
\multicolumn{2}{|c|}{$m_b({\rm GeV})$}
 & 4.8 & 4.9 & 5.0 & 4.8 & 4.9 & 5.0 \\ \hline
 & $t_1^{LO}$ & -1.40-2.29I & -1.87-2.29I & -2.44-2.12I
& -0.68-2.32I & -0.97-2.44I & -1.35-2.66I \\\cline{2-8}\cline{8-8}
$\phi_B^{BHL}$ & $t_1^{NLO}$ 
& 0.01-0.29I & -0.06-0.31I & -0.14-0.30I
& 0.12-0.28I & 0.08-0.32I  & 0.03-0.38I \\\cline{2-8}\cline{8-8}
 & $t_2^{NLO}$ & -0.79 & -0.84 & -0.89 
& -0.86 & -0.91 & -0.97 \\\hline
 & $t_1^{LO}$ & -1.80-2.94I & -2.83-3.45I & -4.63-4.03I
& -0.88-2.88I & -1.47-3.69I & -2.58-5.05I \\\cline{2-8}\cline{8-8}
$\phi_B^{SHB}$ & $t_1^{NLO}$ 
& 0.01-0.37I & -0.08-0.47I & -0.27-0.58I
& 0.15-0.36I & 0.13-0.50I &  0.06-0.72I \\\cline{2-8}\cline{8-8}
 & $t_2^{NLO}$ & -1.63& 2.47 & -4.33 
& -1.76 & -2.69 & -8.13 \\\hline 
 & $t_1^{LO}$ & -1.96-3.24I & -3.24-3.96I & -5.63-4.87I
& -0.96-3.15I & -1.68-4.24I & -3.13-6.15I \\\cline{2-8}\cline{8-8}
$\phi_B^{\delta}$ & $t_1^{NLO}$
& 0.01-0.41I & -0.10-0.53I & -0.33-0.70I
& 0.16-0.40I & 0.15-0.56I & 0.80-0.87I \\\cline{2-8}\cline{8-8}
 & $t_2^{NLO}$ & -0.95 & -1.57 & -3.02
&-1.03 & -1.72 & -3.29 \\\hline
\end{tabular}
\vskip 0.5cm
Table 2. Decay amplitudes in units of $10^{-8} {\rm GeV}$.
\end{center}

\vskip 1cm
\begin{center}
\begin{tabular}{|c|c|c|c|c|c|c|c|} \hline
\multicolumn{2}{|c|}{} & 
\multicolumn{3}{|c|}{$\phi_{k^*}^{GH}$} & 
\multicolumn{3}{|c|}{$\phi_{k^*}^{as}$} \\ \hline
\multicolumn{2}{|c|}{$m_b({\rm GeV})$}
 & 4.8 & 4.9 & 5.0 & 4.8 & 4.9 & 5.0 \\ \hline
 & $Br^{LO}$ & 0.60 & 0.73 & 0.87
& 0.46 & 0.58 & 0.74 \\\cline{2-8}\cline{8-8}
$\phi_B^{BHL}$ & $Br^{Full}$ 
& 0.95 & 1.20 & 1.50
& 0.70 & 0.91 & 1.21 \\\cline{2-8}\cline{8-8}
 & $\frac{Br^{Full}-Br^{LO}}{Br^{Full}}$ & 37\% & 39\% & 42\% 
& 34\% & 36\% & 39\% \\\hline
 & $Br^{LO}$ & 0.99 & 1.66 & 3.16
& 0.76 & 1.32 & 2.69 \\\cline{2-8}\cline{8-8}
$\phi_B^{SHB}$ & $Br^{Full}$
& 1.89 & 3.70 & 8.90
& 1.40 & 2.83 &  7.16 \\\cline{2-8}\cline{8-8}
 & $\frac{Br^{Full}-Br^{LO}}{Br^{Full}}$ & 48\% & 53\% & 65\% 
& 46\% & 53\% & 62\% \\\hline 
 & $Br^{LO}$ & 1.19 & 2.19 & 4.65
& 0.90 & 1.74 & 3.96 \\\cline{2-8}\cline{8-8}
$\phi_B^{\delta}$  & $Br^{Full}$
& 1.80 & 3.71 & 9.35
& 1.33 & 2.82 & 7.47 \\\cline{2-8}\cline{8-8}
 & $\frac{Br^{Full}-Br^{LO}}{Br^{Full}}$ & 34\% & 41\% & 50\%
& 32\% & 38\% & 50\% \\\hline
\end{tabular}
\vskip 0.5cm
Table 3. Branching ratio $Br(B\rightarrow K^*\gamma)$
in units of $\times 10^{-5}$.
\end{center}

\newpage
\bigskip
\noindent
{\bf Figure Caption}

\noindent
\begin{description}
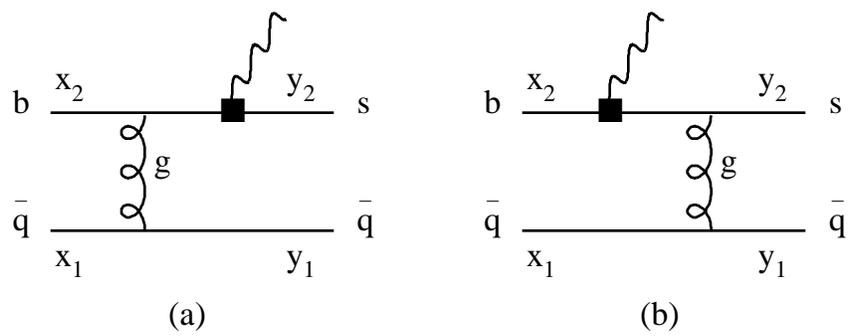
\item{Fig.~1.} The photonic penguin diagram. The square blob represents
the effective vertex.
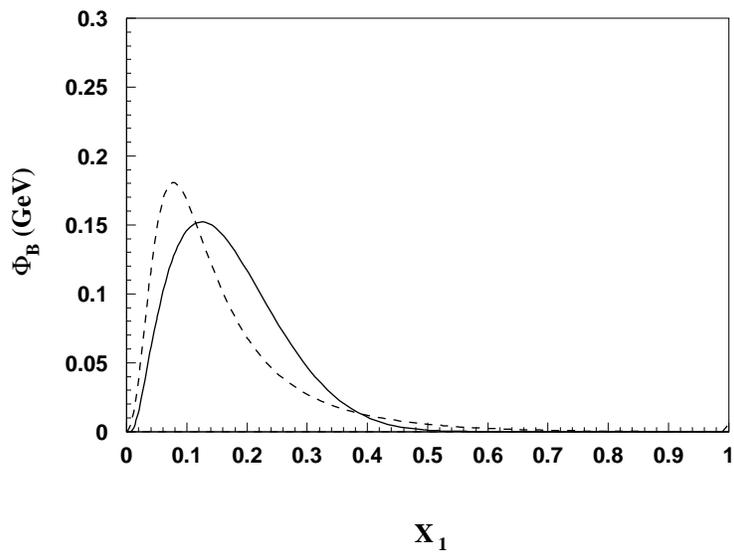
\item{Fig.~2.} The distribution amplitudes of $B$ meson employed in our
calculation: $\phi_B^{BHL}$(the solid curve) with $m_b=4.9$ GeV and
$m_q=0.35$ GeV;
$\phi_B^{SHB}$(the dashed curve) with $\epsilon_B=0.072$;
$\phi_B^{\delta} \sim \delta(x_1-0.072)$ is not plotted in this figure.
\end{description}


\begin{thebibliography}{99}
\bibitem{CLEO}CLEO Collaboration, R. Ammar {\sl et al.} Phys. Rev. 
Lett. {71} (1993) 674.
\bibitem{NSM1}J. Hewitt, Phys. Rev. Lett. {70} (1993) 1045;
V. Barger, M. Berger, and R. Phillips, {\sl ibid.} {70} (1993) 1368.
\bibitem{NSM2}R. Barbieri and G.F. Diudice, Phys. Lett. B {309} 
(1993) 86; R. Garisto and J.N. Ng, {\sl ibid.} {315} (1993) 372.
\bibitem{Brod}A. Szczepaniak, E.M. Henley, and S.J. Brodsky, 
Phys. Lett. B {243} (1990) 287.
\bibitem{Carlson}see {\sl e.g.} C.E. Carlson and J. Milana,
Phys. Lett. B {301} (1993) 237; C.E. Carlson and J. Milana,
Phys. Rev. D {49} (1994) 5908.
\bibitem{CarlsonK}C.E. Carlson and J. Milana, Phys. Rev. D {51}
(1995) 4950.
\bibitem{Heff1}B. Grinstein, R. Springer, and M.B. Wise, Nucl. Phys.  
{B339} (1990) 269; M. Misiak, Phys. Lett B {269} (1991) 161.
\bibitem{Heff2}R. Grigjanus {\sl et al.}, Phys. Lett. B {213}
(1988) 355; G. Cella {\sl et al.}, {\sl ibid.} {248} (1990) 181.
\bibitem{Heff3}M. Ciuchini, E. Franco, G. Martinelli, L. Reina, and
L. Silvestrini, Phys. Lett. B {316} (1993) 127; G. Cella, G. Curci,
G. Giudice, and A. Vicere, {\sl ibid}. {325} (1994) 227.
\bibitem{Zhang}D.X. Zhang, private communication.
\bibitem{fB}C.W. Bernard, J.N. Labrenz, and A. Soni, Phys. Rev. D {49}
(1994) 2536.
\bibitem{fK}P. Colangelo, C. A. Dominguez, G. Nardulli, and N. Paver,
Phys. Lett. B {317} (1993) 183.
\bibitem{BHL}
S.J. Brodsky, T.Huang, and G.P. Lepage, in {\sl Particles
and Fields-2}, Proceedings of the Banff Summer Institute, Banff, Alberta,
1981, edited by A.Z. Capri and A.N. Kamal (Plenum, New York, 1983), p. 143.
G.P. Lepage, S.J. Brodsky, T. Huang, and P.B. Mackenize, {\sl ibid.}, p. 83;
T. Huang, in {\sl Proceedings of XX-th International Conference on High
Energy Physics}, Madison, Wisconsin, 1980, edited by L. Durand and 
L.G. Pondorm, AIP Conf. Proc. No. 69 (AIP, New York, 1981), p. 1000.
\bibitem{Guo}X.H. Guo and T. Huang, Phys. Rev. D {43} (1991) 2931. 
\bibitem{Xiang}X.D. Xiang, X.N. Wang, and T. Huang, Commun. Theor.
Phys. {5} (1986) 117; T. Huang, X.D. Xiang, and X.N. Wang,
Chin. Phys. Lett. {2} (1985) 67.
\bibitem{ZD}Z. Dziembowski and L. Mankiewicz, Phys. Rev. Lett. {58}
(1987) 2175.
\end{thebibliography}
\end{document}